\documentclass{appolb}
\usepackage{epsfig}
\usepackage{color}
\usepackage{subfigure}    
\begin{document}
\title{Strange matter in core-collapse supernovae
\thanks{Presented at the "Strangeness in Quark Matter" conference, 18-24 September 2011, Polish Academy of Arts and Sciences, Cracow, Poland}}
\author{I. Sagert$^1$, T. Fischer$^2$, M. Hempel$^3$, G. Pagliara$^4$, J. Schaffner-Bielich$^5$,
F.-K. Thielemann$^3$, M. Liebend\"orfer$^3$
\address{$^1$Department of Physics and Astronomy, Michigan State University, East Lansing, 48823 MI, USA\\
$^2$GSI Helmholtzzentrum f\"ur Schwerionenforschung GmbH, 64289 Darmstadt, Germany\\
$^3$Department of Physics, University of Basel, 4056 Basel, Switzerland\\
$^4$Dip.~di Fisica dell'Universit\`a di Ferrara and INFN Sez.~di Ferrara, I-44100 Ferrara, Italy\\
$^5$Institut f\"ur Theoretische Physik, Ruprecht-Karls-Universit\"at, D-69120 Heidelberg, Germany}}
\maketitle
\begin{abstract}
We discuss the possible impact of strange quark matter on the evolution of core-collapse supernovae with emphasis on low 
critical densities for the quark-hadron phase transition. For such cases the hot proto-neutron star can collapse to a more 
compact hybrid star configuration hundreds of milliseconds after core-bounce. The collapse triggers the formation of a second 
shock wave. The latter leads to a successful supernova explosion and leaves an imprint on the neutrino signal. These dynamical 
features are discussed with respect to their compatibility with recent neutron star mass measurements which indicate a stiff high 
density nuclear matter equation of state.
\end{abstract}
\section{Introduction}
A core-collapse supernova (SN) explosion marks the disruption of a massive star by an energetic shock wave followed by the 
formation of a neutron star or a black hole. In the first hundreds of milliseconds (ms) of a supernova, temperatures in the range
of tens of MeV, densities beyond nuclear matter saturation density $n_0 \sim 0.145\:$fm$^{-3}$ and proton fractions $Y_p \leq 0.3$ 
are reached. With such properties, supernovae (SNe) are astrophysical laboratories for dense nuclear matter which, 
in the phase diagram of strongly interacting matter, possess overlap regions with heavy-ion experiments, such as in the future 
FAIR facility at GSI, Darmstadt (Germany) and the NICA facility at the JINR in Dubna (Russia). The modeling of core-collapse 
SNe represents a great computational challenge. It requires as input amongst others a nuclear matter equation of state (EoS), 
which  provides information about the thermodynamic properties and compositions of matter for a large range of baryon 
number densities $n_b$, temperatures $T$, and isospin states, characterized by $Y_p$. Being discussed to populate neutron 
star interiors, hyperons and quark matter can also be included in SN equations of state (EoSs). At present, both components are 
tested on their impacts in SN simulations. Hereby, the applied quark and hyperon EoSs should be compatible with
observed neutron star properties, e.g. pulsar masses. As we will argue in the next section, the latter indicate a stiff nuclear matter 
EoS at high density. 
\newline
In the following we will give a short summary on recent pulsar mass measurements and their implications for quark matter 
in neutron star interiors. We will proceed with an overview of quark matter studies in core-collapse SNe. In the 
last section, we will focus on the impact of low density quark-hadron phase transitions on the gravitational collapse of light 
and intermediate mass progenitor stars.
\section{Strange quark matter in massive neutron stars}
Neutron star masses can be deduced for pulsars in binary systems, if effects from general relativity such as the advance of 
periastron $\dot{\omega}$ and Shapiro delay are observable in the pulsar signal \cite{Lorimer08}. For a long time the Hulse-Taylor 
pulsar PSR B1913+16 had the highest precisely known mass of $M = 1.4414 \pm 0.0002\:$M$_\odot$, with most pulsars 
clustering around this value \cite{Thorsett99}. However, the observation of $\dot{\omega}$ and Shapiro delay in the low mass 
X-ray binary J1903-0327 \cite{Champion08} allowed the determination of the mass for the corresponding millisecond pulsar to 
$M = 1.667 \pm 0.021\:$M$_\odot$. In 2010, an even higher pulsar mass was obtained for the millisecond pulsar PSR J1614-2230. 
A massive binary partner and the large inclination angle of the system allow the measurement of the Shapiro delay to a high 
accuracy and reveal a pulsar mass of $M = 1.97 \pm 0.04$M$_\odot$ \cite{Demorest10}. At present, this value represents the 
highest robust mass ever measured for a compact star and poses tight constraints on the nuclear matter EoS restricting it to 
stiff models. Hybrid and strange stars are only compatible with such a high mass if stiffening effects from the strong interaction 
between the quarks, e.g. in form of the strong interaction coupling constant $\alpha_s$ and/or color superconductivity - 
have a large impact on the quark matter EoS (see \cite{Weissenborn11} and references therein). 
\section{Strange quark matter in core-collapse supernovae}
The dynamical effects of phase transitions on neutron stars and core-collapse SNe were first discussed by Migdal \cite{Migdal79} 
in 1979, and later studied by Takahara and Sato \cite{Takahara86} as well as Brown \cite{Brown88}. In the early 90s, 
Gentile et al. \cite{Gentile93} performed general relativistic hydrodynamic core-collapse SN simulations including a phase 
transition to strange quark matter for critical densities of $\sim (2-3)\: n_0$. The authors tested different setups of the 
quark-hadron mixed phase, finding the formation of two shock waves. The first shock wave is caused by the usual stiffening 
of the hadronic EoS for $n_b > n_0$, which halts and reverts the infall of matter in the center of the collapsing iron core 
(core-bounce). The second wave is formed due to the softening of the EoS in the quark-hadron mixed phase together with the 
subsequent stiffening in pure quark matter. As neutrino transport was not included in the calculations, the dynamics of the 
shock waves could only be followed for a few ms. Simulations including neutrino transport were performed by e.g. Nakazato 
et al. \cite{Nakazato08, Nakazato10}. The authors applied the hadronic SN Shen EoS \cite{Shen98} extending it by the inclusion 
of quark matter at higher densities as well as thermal pions. For the quark matter EoS, Nakazato et al. chose the simple bag 
model \cite{Chodos74b} with a bag constant of $B^{1/4} \sim 209\:$MeV. The quark-hadron phase transition was modeled 
by a Gibbs construction \cite{Glendenning92}. The studied progenitor stars had masses of $\geq 40\:$M$_\odot$, whereas 
the gravitational collapse of such massive stars usually ends with the production of a black hole \cite{Heger03}. Due to the 
softer hybrid EoS, the onset of quark matter accelerates the black hole formations by tens to hundreds of ms in comparison 
to simulations with the original Shen EoS. While the earlier collapse results in a shortening of the neutrino emission, the softening
of the EoS influences the neutrino spectra. Both effects can be used as indicators for a quark matter phase transition. 
However, as hyperons and different nuclear interactions \cite{Sumiyoshi10, Hempel11} can have similar imprints on the neutrino 
signal, further studies are required. 
\newline
In this work we focus on quark matter phase transitions with low critical densities around $n_0$ for SN conditions. 
It was shown by Drago et al. \cite{Drago99} and later argued in Sagert et al. \cite{Sagert09} and Fischer et al. \cite{Fischer11} 
that quark models with such low $n_{crit}$ for SN conditions can be compatible with higher transition densities in heavy ion 
collisions, caused by the larger $Y_p$ in the latter and different strangeness production mechanisms.
\section{Early quark-hadron transition in core-collapse supernova}
In our study we implement a quark-hadron phase transition into the Shen EoS via the Gibbs construction. For the quark 
matter EoS, we apply a quark bag model which is extended by first order corrections in $\alpha_s$. The thermodynamic grand potential is 
given by \cite{Fischer11}: 
\begin{equation}
\Omega_{QM}= \sum_{i=u,d,s} \left[\Omega_i + \frac{\alpha_s}{\pi} \left( T^2 \mu_i^2 + \frac{\mu_i^2}{2 \pi^2} \right) + \frac{35 \pi}{126} T^4 \alpha_s \right] + B .
\label{bag_eos}
\end{equation}
Hereby, $\Omega_i$ are the fermi contributions for the up, down, and strange quarks. First order corrections from $\alpha_s$ 
are given by the second and third terms, while $B$ is the bag constant which represents all non-perturbative effects of the strong 
interaction. The quark masses are chosen to $m_s=100\:$MeV for strange quarks and $m_u=m_d=0\:$MeV for up and down quarks. 
In the very simple bag model, first order corrections from $\alpha_s$ are not present. As a consequence, the corresponding quark 
EoS is very soft. The maximum masses of the resulting hybrid stars are low and do not fulfill the mass constraint of PSR  
J1614-2230 \cite{Weissenborn11}. The inclusion of $\alpha_s$ correction terms stiffen the quark EoS and can increase the hybrid 
star maximum masses up to $\geq 2\:$M$_\odot$ \cite{Weissenborn11}. 
\begin{figure}
\begin{center}
\includegraphics[width=0.45\textwidth, angle=270]{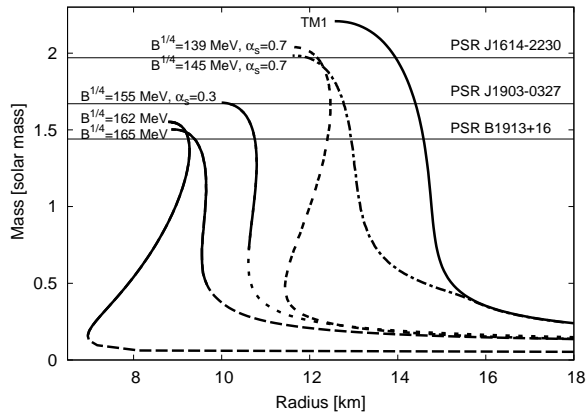}
\caption{Hybrid star mass radius relations for the quark bag model in eq.(\ref{bag_eos}) and the hadronic EoS TM1. 
Solid lines are stars with a pure quark matter core, while dashed lines indicate mixed phase cores. The horizontal
lines show pulsar mass measurements discussed in the text.}
\label{mr_plot1}
\end{center}
\end{figure}
For low critical densities, the resulting hybrid stars have an extended quark-hadron mixed phase in their interior and only for 
sufficiently stiff hadronic EoSs possess a pure quark matter core \cite{Weissenborn11}. Figure \ref{mr_plot1} shows the mass-radius 
relations for different parameter sets in eq.(\ref{bag_eos}). Hadronic matter is described by the relativistic mean field TM1 model 
\cite{Sugahara94} which is also the basis for the Shen EoS. The three parameter sets with $\alpha_s=0$ and $\alpha_s=0.3$, shown 
in Fig.~\ref{mr_plot1}, have been tested on their impact in one dimensional SN simulations based on general relativistic radiation 
hydrodynamics and three flavor Boltzmann neutrino transport \cite{Liebendoerfer04}. The simulations show that due to the low 
critical densities, the quark-hadron mixed phase sets in already at core-bounce, when matter in the center of the collapsing iron 
core reaches densities around $n_0$. However, as the quark fraction is small and $Y_p \sim 0.3$, the hybrid EoS is very similar 
to the one of hadronic matter \cite{Fischer11}, and the dynamics proceed like in a normal core-collapse SN for the first $(200-400)\:$ms. 
The core-bounce launches a hydrodynamic shock wave which starts to propagates outwards. On its way, it loses energy due to 
the disintegration of infalling heavy nuclei and production of neutrinos. As the shock wave moves across the neutrinospheres, 
the neutrinos are emitted in a burst dominated by electron neutrinos $\nu_e$. This energy loss turns the expanding shock into a 
standing accretion shock (SAS). As shock-heated matter continues to be accreted through the SAS on the surface of the proto neutron 
star (PNS), the density and temperature in its interior rise and a growing volume of the PNS enters the mixed phase. The quark fraction in 
the mixed phase increases which leads to a softening of the EoS. This, together with the growing gravitational mass of the PNS eventually 
triggers its collapse to a more compact hybrid star configuration. Similar to the study of Gentile et al., a second shock wave forms and 
propagates outwards. Shock heating of the infalling neutron rich matter leads to the production of $\bar{\nu}_e$, as well as $\nu_{\mu / \tau }$, 
and $\bar{\nu}_{\mu / \tau }$, which are released in a second burst as the second shock wave propagates across the neutrinospheres.
\begin{figure}
\begin{center}
\includegraphics[width=0.42\textwidth, angle=270]{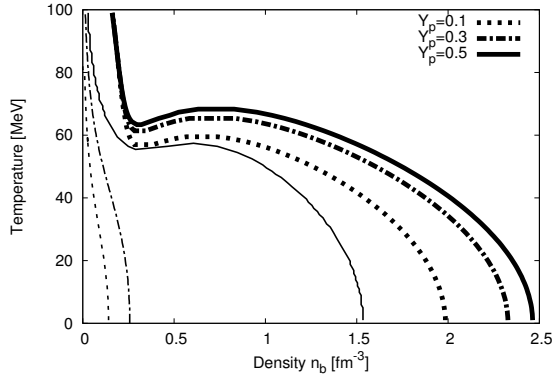}
\caption{Phase diagram for transitions from hadronic matter (TM1) to strange quark matter at $Y_p=0.1,0.3$, and $0.5$.
The quark matter parameters are $B^{1/4}=145\:$MeV and $\alpha_s=0.7$. Thin lines mark the onset of the quark-hadron mixed phase, 
thick lines give the transition to pure quark matter.}
\label{phase_diag145_as07}
\end{center}
\end{figure}
For all three parameter sets the second shock wave eventually leads to a SN explosion, whereas the explosion energy and the time 
difference between the first and second neutrino bursts depend on $n_{crit}$ and the model of the progenitor star \cite{Sagert09,Fischer11}. 
To test the effects of a stiffer quark EoS on the SN dynamics we chose $B^{1/4}=145\:$MeV and $\alpha_s=0.7$. As can be seen from 
Fig.~\ref{mr_plot1}, the corresponding hybrid star maximum mass is $\sim 1.97\:$M$_\odot$ and thereby compatible PSR J1614-2230. 
Fig.~\ref{phase_diag145_as07} shows the phase diagram for different values of $Y_p$ (for details of phase diagram calculations 
see \cite{Fischer11}). The critical densities for the onset of the mixed phase for $Y_p = 0.3$ are around $n_{crit} < 1.5\:n_0$, however, 
due to the similar stiffness of the quark and hadron EoSs, the mixed phase extends up to $\geq 10\: n_0$ for 
low T. As the temperature rises above $60\:$MeV, the mixed phase experiences a significant reduction and the critical densities 
become low. However, for such high values of $T$, the inclusion of pions and hadron resonances becomes important and would most 
likely change the shape of the phase diagram. 
\begin{figure}
\begin{center}
\includegraphics[width=0.45\textwidth, angle=270]{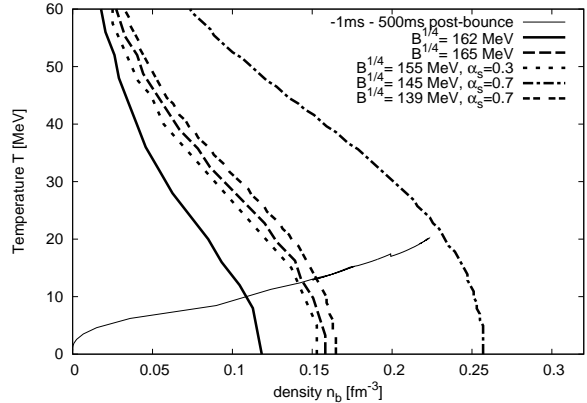}
\caption{Onset of the mixed phase for different quark EoS parameter sets and $Y_p=0.3$. The thin solid line
show the temperatures and densities which are reached in the center of the PNS shortly before and during the first $500\:$ms after 
core-bounce.}
\label{phase_diag_as07_a}
\end{center}
\end{figure}
Similar to the soft bag EoS parameter sets, we tested the new hybrid EoS in one dimensional SN simulations 
for the collapse of a $15\:$M$_\odot$ and a $30\:$M$_\odot$ progenitor. In both cases we find that quark matter appears very 
late - around $1\:$s after core bounce and does not influence the SN dynamics \cite{Sagert12}. The reason for this outcome 
can be seen in Fig.~\ref{phase_diag_as07_a} which shows the onset of the mixed phases for the discussed quark EoSs. 
Though the critical densities for $B^{1/4}=145\:$MeV and $\alpha_s=0.7$ are low, it can be seen that they 
are higher in comparison to the soft EoSs. In addition we plot the densities and temperatures in the center of the PNS 
shortly before and during the first $500\:$ms after the core-bounce. While it can be seen that for the soft models, matter in the PNS 
enters the quark-hadron mixed phase, the critical densities for the new parameter set are too high. 
However, the reduction of the bag constant to $B^{1/4}=139\:$MeV leads to sufficiently low values of $n_{crit}$ as is illustrated in 
Fig.~\ref{phase_diag_as07_a} and also fulfills the maximum mass constraint of PSR J1614-2230 (see Fig.~\ref{mr_plot1}).
This parameter set is currently tested in one dimensional SN simulations and will give further insights in the role and detectability 
of a quark matter phase transition in core-collapse SNe.
\section{Summary}
The onset of strangeness in core-collapse supernovae (SNe) can be studied by implementing hyperons and strange quark matte in SN 
equations of state (EoSs). Hereby, the chosen parameter sets must fulfill restrictions from the recent finding of a two solar mass pulsar 
PSR J1614-2230, which is only compatible with a stiff high density nuclear matter EoS. Within currently applied quark EoS models, 
high critical densities were shown to shorten the time to black hole formation in the gravitational collapse of massive progenitor stars. 
For the onset of quark matter with a soft EoS at critical densities $n_{crit}$ around nuclear matter saturation density $n_0\sim 0.145$fm$^{-3}$, 
the proto neutron star collapses to a more compact hybrid star configuration within hundreds of ms after core-bounce. This launches a 
shock wave which leads to the SN explosion and releases a neutrino burst, dominated by $\bar{\nu}_e$. The properties of the neutrino 
burst are dependent on $n_{crit}$ as well as the model of the progenitor star. A first study with a stiff quark EoS shows no 
impacts on the SN dynamics as the critical densities are too high to be reached in the early post-bounce phase. However, studies are 
on the way, in which we apply a stiff quark EoS parameter set where SN matter enters the quark-hadron mixed phase at $n_{crit} \sim n_0$.\\
\newline
\textit{Acknowledgements}\\
The project was funded by the Swiss National Science Foundation (SNF) under project numbers PP00P2 - 124879/1, 200020 - 122287. 
T.F is support by HIC for FAIR and by the SNF under project~no.~PBBSP2-133378. The work of G.P. is supported by 
the DFG under Grant No. PA 1780/2-1 and J.S.-B. is supported by the DFG through the Heidelberg Graduate School of Fundamental 
Physics. M. H. is supported by the SNF under project number no. 200020-132816/1. M. H. is also grateful for 
participating in the EuroGENESIS collaborative research program of the 
European Science Foundation (ESF) and the ENSAR/THEXO project. I.S. is supported by the AvH foundation via a Feodor Lynen fellowship 
and wishes to acknowledge the HPCC of MSU and the iCER. The authors are additionally supported by CompStar, a research networking 
program of the ESF, and the scopes project funded by the SNF grant. no. IB7320-110996/1.
\bibliographystyle{nar}

\begin{thebibliography}{10}

\bibitem{Lorimer08}
{Lorimer}, D.~R. (2008)
{\em Living Reviews in Relativity} {\bf 11}, 8--+.

\bibitem{Thorsett99}
{Thorsett}, S.~E. and {Chakrabarty}, D. February 1999
{\em Astrophys. Journal} {\bf 512}, 288--299.

\bibitem{Champion08}
{Champion et al.}, D.~J. (2008)
{\em Science} {\bf 320}, 1309--.

\bibitem{Demorest10}
{Demorest}, P.~B., {Pennucci}, T., {Ransom}, S.~M., {Roberts}, M.~S.~E., and
  {Hessels}, J.~W.~T. (2010)
{\em Nature} {\bf 467}, 1.

\bibitem{Weissenborn11}
{Weissenborn}, S., {Sagert}, I., {Pagliara}, G., {Hempel}, M., and
  {Schaffner-Bielich}, J. (2011)
{\em Astrophys. Journal Lett.} {\bf 740}, L14.

\bibitem{Migdal79}
{Migdal}, A.~B., {Chernoutsan}, A.~I., and {Mishustin}, I.~N. (1979)
{\em Phys. Lett. B} {\bf 83}, 158--160.

\bibitem{Takahara86}
{Takahara}, M. and {Sato}, K. (1986)
{\em Astrophys. \& Space Science} {\bf 119}, 45--49.

\bibitem{Brown88}
{Brown}, G.~E. (1988)
{\em Zeitschrift fur Physik C} {\bf 38}, 291--301.

\bibitem{Gentile93}
{Gentile}, N.~A., {Aufderheide}, M.~B., {Mathews}, G.~J., {Swesty}, F.~D., and
  {Fuller}, G.~M. (1993)
{\em Astrophys. Journal} {\bf 414}, 701--711.

\bibitem{Nakazato08}
Nakazato, K., Sumiyoshi, K., and Yamada, S. (2008)
{\em Phys. Rev. D} {\bf 77(10)}, 103006.

\bibitem{Nakazato10}
{Nakazato}, K., {Sumiyoshi}, K., and {Yamada}, S. (2010)
{\em Astrophys. Journal} {\bf 721}, 1284--1294.

\bibitem{Shen98}
{Shen}, H., {Toki}, H., {Oyamatsu}, K., and {Sumiyoshi}, K. (1998)
{\em Progress of Theoretical Physics} {\bf 100}, 1013--1031.

\bibitem{Chodos74b}
{Chodos}, A., {Jaffe}, R.~L., {Johnson}, K., and {Thorn}, C.~B. (1974)
{\em Phys. Rev. D} {\bf 10}, 2599--2604.

\bibitem{Glendenning92}
{Glendenning}, N.~K. (1992)
{\em Physical Review D} {\bf 46}, 1274--1287.

\bibitem{Heger03}
{Heger}, A., {Fryer}, C.~L., {Woosley}, S.~E., {Langer}, N., and {Hartmann},
  D.~H. (2003)
{\em Astrophys. Journal} {\bf 591}, 288--300.

\bibitem{Sumiyoshi10}
{Sumiyoshi}, K., {Nakazato}, K., {Ishizuka}, C., {Ohnishi}, A., {Yamada}, S.,
  and {Suzuki}, H. (2010)
{\em Nucl. Phys. A} {\bf 835}, 295--302.

\bibitem{Hempel11}
{Hempel}, M., {Fischer}, T., {Schaffner-Bielich}, J., and {Liebend{\"o}rfer},
  M. (2011)
{\em ArXiv e-prints}.

\bibitem{Drago99}
{Drago}, A. and {Tambini}, U. (1999)
{\em Journal of Phys. G} {\bf 25}, 971--979.

\bibitem{Sagert09}
{Sagert}, I., {Fischer}, T., {Hempel}, M., {Pagliara}, G., {Schaffner-Bielich},
  J., {Mezzacappa}, A., {Thielemann}, F.-K., and {Liebend{\"o}rfer}, M. (2009)
{\em Phys. Rev. Lett.} {\bf 102(8)}, 081101.

\bibitem{Fischer11}
{Fischer}, T., {Sagert}, I., {Pagliara}, G., {Hempel}, M., {Schaffner-Bielich},
  J., {Rauscher}, T., {Thielemann}, F.-K., {K{\"a}ppeli}, R.,
  {Mart{\'{\i}}nez-Pinedo}, G., and {Liebend{\"o}rfer}, M. (2011)
{\em Astrophys. Journal, Suppl.} {\bf 194}, 39.

\bibitem{Sugahara94}
{Sugahara}, Y. and {Toki}, H. October 1994
{\em Nuclear Physics A} {\bf 579}, 557--572.

\bibitem{Liebendoerfer04}
{Liebend{\"o}rfer}, M., {Messer}, O.~E.~B., {Mezzacappa}, A., {Bruenn}, S.~W.,
  {Cardall}, C.~Y., and {Thielemann}, F. (2004)
{\em Astrophys. Journal, Suppl.} {\bf 150}, 263--316.

\bibitem{Sagert12}
In preparation.

\end{thebibliography}

\end{document}